# Low optical loss electrical isolation for multi-section monolithic GaSb-based photonic circuits

**Md Ajwaad Zaman Quashef [(a)], Nouman Zia, Jukka Viheriälä, and Mircea Guina**

*Optoelectronics Research Centre, Physics Unit, Tampere University, Korkeakoulunkatu 3, 33720 Tampere, Finland*

(a) Author to whom correspondence should be addressed: md.quashef@tuni.fi

## Abstract

Monolithic photonic integrated circuits (PICs) platforms exploiting III-V materials combine passive and active waveguide structures in multi-section optoelectronic device architectures. Their operation requires high electrical isolation between adjacent functional sections without compromising the optical signal. This fundamental requirement is addressed for GaSb-based waveguides, which are known to exhibit high conductivity of p-type layers reducing the electrical isolation capability. To this end, a co-designed electrical–optical isolation strategy based on using deeply etched strip waveguides combined with adiabatic ridge-to-strip waveguide tapers in GaSb-based multiple-quantum-well heterostructures is proposed. While deep etching alone enables isolation resistances of up to 40 kΩ, it severely degrades optical propagation. By introducing optimized adiabatic tapers, we demonstrate good optical performance as single-mode continuous-wave lasing in a two-section device with integrated absorber, while maintaining an isolation resistance of 17.3 kΩ; this corresponds to an approximately 17-fold improvement over previously reported GaSb two-section devices. The approach establishes a critical building block for the development of monolithic GaSb-based PICs operating above 2 μm.

Driven by the sustained demands of the telecommunications industry, the monolithic photonic integrated circuits (PICs) platform based on InP has reached a high maturity level [1], [2]. It enables dense integration of diverse optical functions, typically operating in the 1.3 and 1.55 µm wavelength regions, with disruptive effects in datacom [3] and more recently 3D sensing [4] applications. Still, the success of InP PICs has been marginal in other application domains, largely because of limited wavelength coverage. An essential step in this direction is achieving improved wavelength versatility, particularly in the mid-infrared (MIDIR) part of the optical spectrum extended beyond 2 µm. This spectral window is rich in "molecular fingerprints" and is highly relevant for optical sensing, enabling development of practical integrated sensors for environmental gases [5] and biomedical markers [6].

To this end, the GaSb material system has enabled the development of key optoelectronic components operating in MIDIR [7], [8] and more recently sparked the interest in developing hybrid GaSb/SiPh PICs [9], [10], [11]. Yet, development of a comprehensive monolithic GaSb integration platform is still a technological gap which requires foremost the ability to cointegrate multi-section active and passive photonic devices on a single GaSb PIC [12], [13]. To this end, the fundamental functionality to be ensured is the ability to independently bias adjacent components, minimizing electrical crosstalk and optical losses; yet achieving good electrical isolation is a complex fabrication step. For mature InP and emerging GaAs platforms this is regularly achieved by controlling semiconductor resistivity locally via mesa, dielectric or proton implantation techniques [14], [15], [16].

High electrical isolation not only prevents leakage or cross-talk into adjacent components or the substrate but also helps to minimize current spreading and laterally confine the optical gain. Suppressing carrier diffusion in the longitudinal direction ensures that adjacent unpumped sections remain highly absorptive, which is critical for accurately restricting the optical gain and characterizing the internal loss [17]. However, compared to purely electronic ICs, electrical isolation sections in PICs must also be carefully designed to not compromise the optical performance of on-chip photonic devices by inducing high absorption or creating unwanted optical interfaces. This aspect becomes more critical in architectures such as monolithic two-section mode-locked laser diodes (MLLDs), which consist of an integrated forward-biased gain section and a reverse-biased saturable absorber [17]. Without proper electrical isolation, increased reverse biasing causes high lateral leakage currents originating from significantly high p-cladding conductivity in GaSb [18], compromising mode-locking stability, inducing pulse jitter, and worsening device reliability due to the risk of thermal and electric breakdown. Similar design challenges are present in electro-absorption modulators [19] co-integrated with semiconductor optical amplifiers [20], [21] and photodiode arrays [22] where reliable reverse biasing is crucial to collect carriers fast and preserve linearity of the photocurrent response.

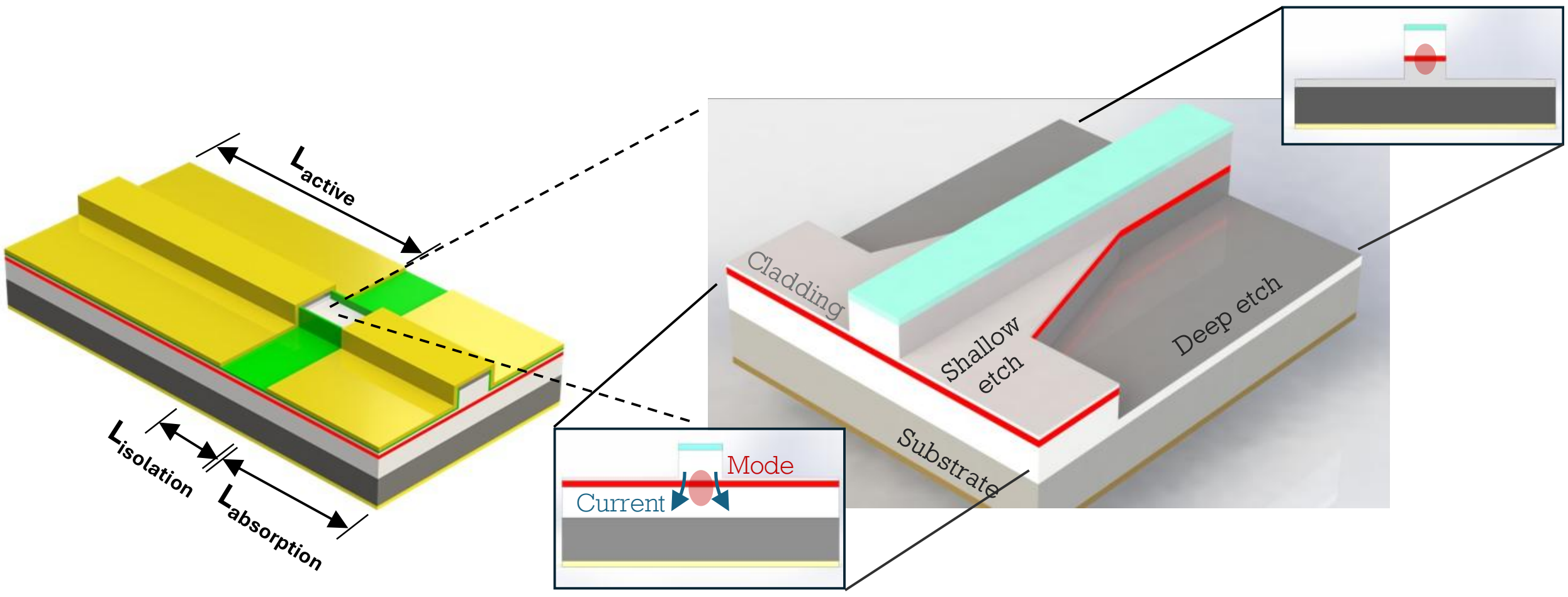

**Fig. 1.** Isometric view of the two-section laser device with its isolation section (left); ridge to strip taper showing the cross sections of ridge (inset, bottom) and strip (inset, top) waveguide (right).

When considering designing electrical isolation sections with low optical loss, a quasi-planar ridge waveguide (RWG) geometry shown in **Fig. 1** is beneficial, ensuring good current confinement and low-loss for single mode waveguides. These features are desirable for laser diode cavities as well as for optical routing between lasers, amplifiers, modulators and other components in PICs [23], [24], [25]. In addition, this geometry also provides easy access to electrical contacts laterally. However, the ridge waveguide is not ideal for electrical isolation because it leaves a wide layer of semiconductor between the etched isolation feature and pn-junction leading to increased lateral leakage current. Implantation with protons to induce crystal damage is a technique typically used to locally increase resistivity and improve current confinement in ridge geometry quantum cascade lasers [26], [27]. Though the technique is commonly used for GaAs and InP-based devices [28], [29], [30], it is found to be fundamentally limited in p-GaSb by the inherently lower bandgap, which facilitates thermal re-excitation of trapped carriers and thus restricts the maximum achievable resistivity. Increasing the implantation dose also causes increased optical absorption [25], [31] and moreover, the method has not been reported for longitudinal electrical isolation in RWG devices. Therefore, the alternative path to make GaSb ridge waveguides suitable for isolation is to etch them deeper and remove most of the semiconductor material that could contribute to electrical conductivity.

Etching of the highly doped p-GaSb cap layer and partially the top p-cladding was reported [32], [33], [34] to achieve inter-section resistances of ~1 kΩ in the design of MLLDs. A more definite approach to reduce current spreading would be to additionally form a strip waveguide (SWG), which is the terminology we use for a deep mesa where the sides are etched completely through the active region and some part of the bottom cladding [35]. However, the tradeoff with optical performance is that strip waveguides are more sensitive to scattering losses caused by interaction of the exposed mode with rough sidewalls [36], [37]. Furthermore, an abrupt transition from a shallower mesa etch in the RWG to a deeper mesa etch in the SWG gives rise to optical losses and reflection arising from mode mismatch at the interface.

This paper addresses the central design conflict between high electrical isolation and adverse optical loss or reflection by employing an adiabatic ridge-to-strip converter monolithically fabricated in GaSb. Deep waveguide etching improves electrical isolation while the ridge to strip waveguide tapering maintains sufficient optical confinement and limits excess scattering sufficiently to support stable CW lasing. The approach has been used in InP and silicon-on-insulator (SOI) photonic integrated platforms to transfer from multi-mode to single mode waveguides or between single mode waveguides of different size/type [38], [39], but here it is utilized for improving the electro-optic functionality. The effectiveness of waveguide etched features and tapered solutions are systematically studied by processing five geometrical variants of two-section GaSb multiple quantum well (MQW) laser diodes and evaluating their continuous wave (CW) laser performance metrics when the absorber section is reverse biased at a large voltage.

First, in order to estimate the mode mismatch at the ridge and strip waveguide interface, mode profiles were simulated using a finite difference eigenmode solver (Ansys Lumerical MODE). The effective index of the fundamental transverse electric (TE) mode in the RWG was determined to be $n_{eff}$=3.581 at the design wavelength of 2.07 μm, and it had a QW confinement factor of 1.75%. A mode match of 81.27% between the RWG and SWG was calculated as an overlap integral which resulted in a ~10% optical loss compared to only RWG propagation, as extracted from eigenmode expansion simulations (EME). The SWG also supports a vertical second order mode ($TE_{01}$) but with a 0.1% QW-confinement resulting in insufficient gain for lasing. Besides, a single mode RWG can be adiabatically coupled to a multimode SWG with almost no coupling to higher order modes [40], especially in a short propagation length. Therefore, the presence of this higher order mode could be neglected. An efficient single mode conversion from a RWG to a SWG depends on several parameters like the length of adiabatic taper $L_{taper}$, width of taper start $W_{taper}$ or taper angle $\theta_{taper}$, given the ridge width $W_{ridge}$, strip width $W_{strip}$ and etch depths at various sections of the device. Finite difference time domain (FDTD) simulations were conducted in Ansys Lumerical FDTD solver to study the influence of the etch depth on the fundamental mode transmission (see **Fig. S1** and **S2** included as supplementary material). Finally, parameter sweeps were conducted using eigenmode expansion in MODE to determine the waveguide width and taper length corresponding to maximal transmission for the fundamental propagation mode. The final optimized parameters are summarized in **Table S1** of the supplementary material**.**

The experimental semiconductor structures were fabricated by solid source molecular beam epitaxy on an n-doped GaSb substrate. Detailed epitaxial design is provided in the supplementary material. For an initial study of electrical isolation, wafer samples were processed employing UV-contact lithography to define multi-section strip waveguides of various widths as illustrated in **Fig. S3** of supplementary material. The sections were separated by trenches having etch depths of 1550 nm and 1800 nm for comparison. Electrical isolation in strip waveguide structures was tested by measuring the current between adjacent metal contacts under applied bias in a four-probe setup. Measured resistances from the IV sweep vs. strip widths are shown in **Fig. 2** for two different isolation etch depths. As predicted, the electrical isolation increased when the strip WG gets narrower and/or isolation etch gets deeper. Moreover, the latter effect is more pronounced at narrower widths where surface states start to play a larger role in current conduction. Whereas active waveguide widths in practice are constrained by single transverse mode operability, the results indicate that slightly deeper isolation etching into the top cladding could be used to enhance the isolation resistance by a factor of at least 3 in narrow laser waveguides.

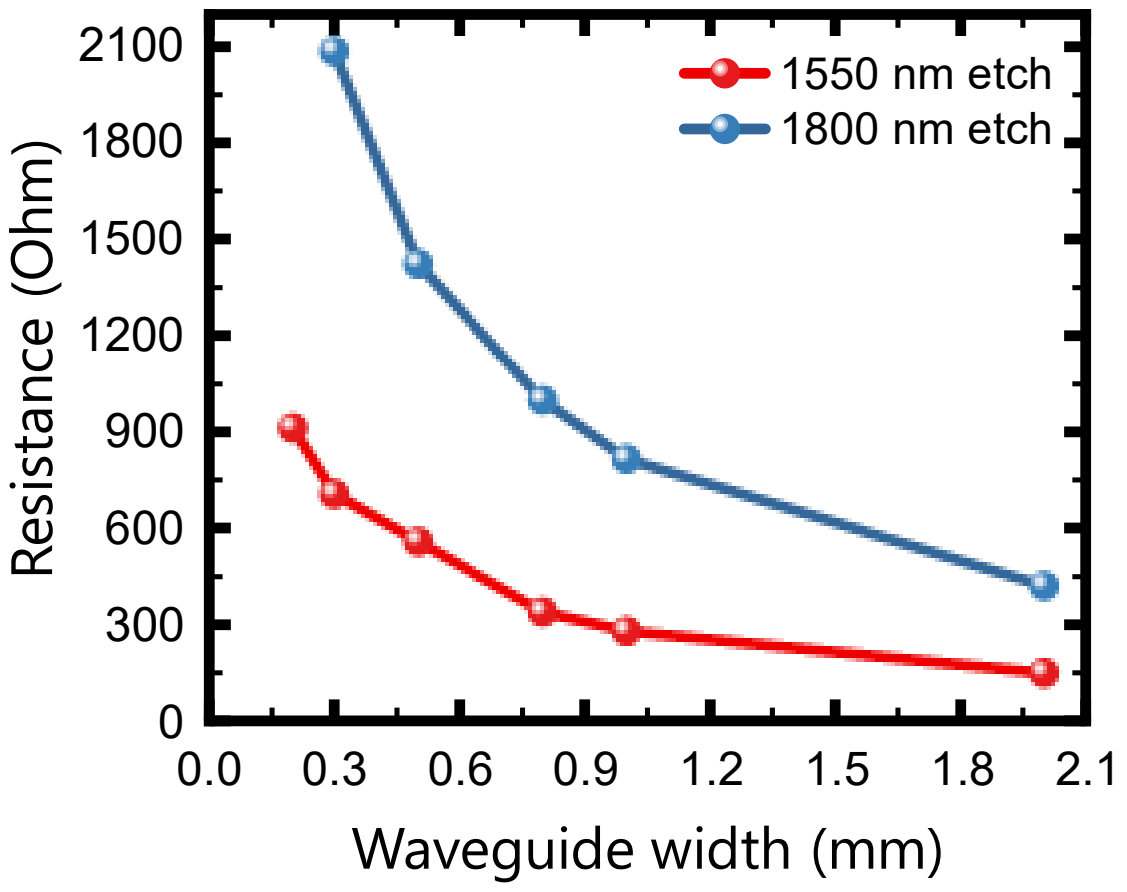

**Fig. 2.** Measured electrical resistance between adjacent sections of test devices vs. strip width for two isolation etch depths. Length of the isolation etch was 40 μm.

For the detailed study, we employ a process to define 3 μm wide RWGs etched 1700 nm through to the lower waveguide layer; for this we used an inductively coupled reactive ion etch system with $Cl_2/N_2$-plasma chemistry. Then to electrically isolate the gain section from the absorber section, a further 1500 nm deep etch was made through the QWs and lower n-waveguide to form a deep mesa in a 620 μm long "isolation trench" region. Two 300 μm long ridge-to-strip tapers were defined in this region as a gradual narrowing of the shallow etch on both sides of the ridge, with a strip waveguide of 20 μm length ($L_{iso}$) called the isolation section. Next, a 150 nm thick $Si_3N_4$ layer was deposited using plasma enhanced chemical vapor deposition (PECVD) technique. The $Si_3N_4$ layer was removed from the top of the waveguide in order to open a path for current injection. Here an isolation etch, also referred to as "leakage prevention etch" (LPE), was defined in the isolation section to remove 1000 nm from the top cladding using a careful combination of $Cl_2/N_2$-plasma dry etch (400 nm) and $H_3PO_4$:$H_2O_2$ wet etch (600 nm), keeping in mind the numerous problems related to deep etching in GaSb [41]. Another lithography step was done to grow a 100 nm thick $SiO_2$ layer on top of the LPE for protection. Afterwards, the p-side metal contact consisting of a Ti/Pt/Au-layer structure was deposited on the $Si_3N_4$ and on the opened area using electron beam evaporation. To allow subsequent cleaving of high-quality facets, the substrate was thinned down to 120 μm thickness and the n-side of the sample was metallized with an annealed Ni/Au/Ge/Au layer stack to produce the n-contact. Each bar containing all the variants was cleaved to 300 μm wide and 3 mm long chips including a 360 μm long absorber on one side. Both facets were left as-cleaved. For testing, the chips were mounted p-side up on Au-plated AlN submounts with an epoxy adhesive containing silver particles and the two sections were wire-bonded to separate pads on the submounts.

To systematically investigate the impact of device geometry, the photomask was designed so that five variants could be processed in parallel from the same wafer as shown in **Fig. 3**. The baseline (Device A) was a two-section RWG device with simple metal isolation, lacking the deep mesa etch and LPE, as well as any waveguide tapers. Device variant B had the leakage prevention etch in the isolation section but no deep mesa etching. Device C included the LPE in the isolation section as well as the deep mesa etch throughout the entire device length. Device D was the co-designed RWG device with waveguide tapering into the deeply etched SWG in the isolation section, which included also the LPE. Finally, Device E replicated Device D but with a longer isolation section ($L_{iso}$=50 μm vs. 20 μm).

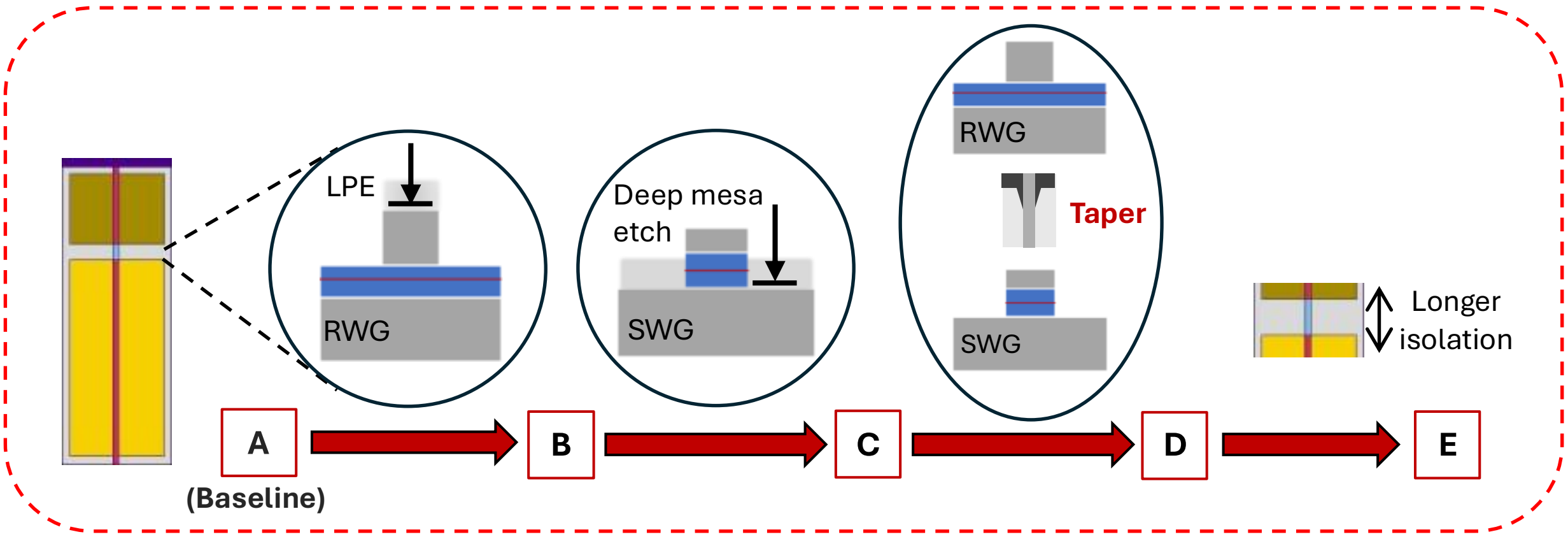

**Fig. 3.** Diagram showing the difference in features between design variants (not to scale)

Mounted laser devices were placed on a Cu heatsink and tested in the CW regime at room temperature (RT). The gain section was biased with a forward direct current while the absorber section was reverse-biased using a custom PCB driven by separate channels from a Keithley 2604B source-measurement unit (SMU). Forward bias current sweeps were carried out for all device variants and the current sunk (hence -ve sign) to the voltage source of the SMU (connected to the absorber section) was measured at different reverse biases. An equivalent circuit model was analyzed to properly extract the isolation resistance $R_{iso}$ as the forward bias voltage of the gain section divided by the leakage current at a sub-lasing threshold forward current (=100 mA was chosen) with the absorber section biased at 0 V. **Fig. 4(a)** highlights the consequences of the poor electrical isolation in the baseline device characterized by a low $R_{iso}$ of 121 Ω, in contrast to **Fig. 4(b)** showing a ~143 times larger $R_{iso}$ of 17.3 kΩ measured in the optimized device (D) having the additional etching steps. The impact of the better isolation is that leakage currents mainly arise from carrier generation in the absorber and show less sensitivity to the magnitude of reverse biasing. Their values were limited to <1 mA (vs. up to 78 mA in the baseline device) in spite of large voltages (up to -5 V) in the absorber. These high currents contributed to concentrated heating effects in the narrow isolation section and compromised the baseline device's reliability, especially when operated as a CW laser. In device B, despite the isolation etch of 1000 nm, the lower-than-expected improvement of $R_{iso}$ (=572 Ω) warranted electrical characterization of the p-cladding. Measurement of the sheet resistance by patterning contacts in the Transfer Length Method (TLM) yielded an average value of 313 Ω/sq for the 1 µm thick layer, following which a 20 µm long and 3 µm wide RWG structure with 200 nm thick p-cladding was projected to have an inter-section resistance of ~10 kΩ. However, it can be expected that conducting surface states, an elemental Sb layer [42], [43], and/or etch residue, which are exposed due to multiple RWG etching steps- all accommodate large surface currents and reduce the measured isolation. The largest $R_{iso}$=39.57 kΩ was measured in device C, while a lower value of 17.3 kΩ was obtained in device D having the deep etch only in the isolation section. It is evident from this that lateral pathways for current along the entire device length can contribute to leakage and may be prevented by deep etching throughout if one targets only the best isolation.

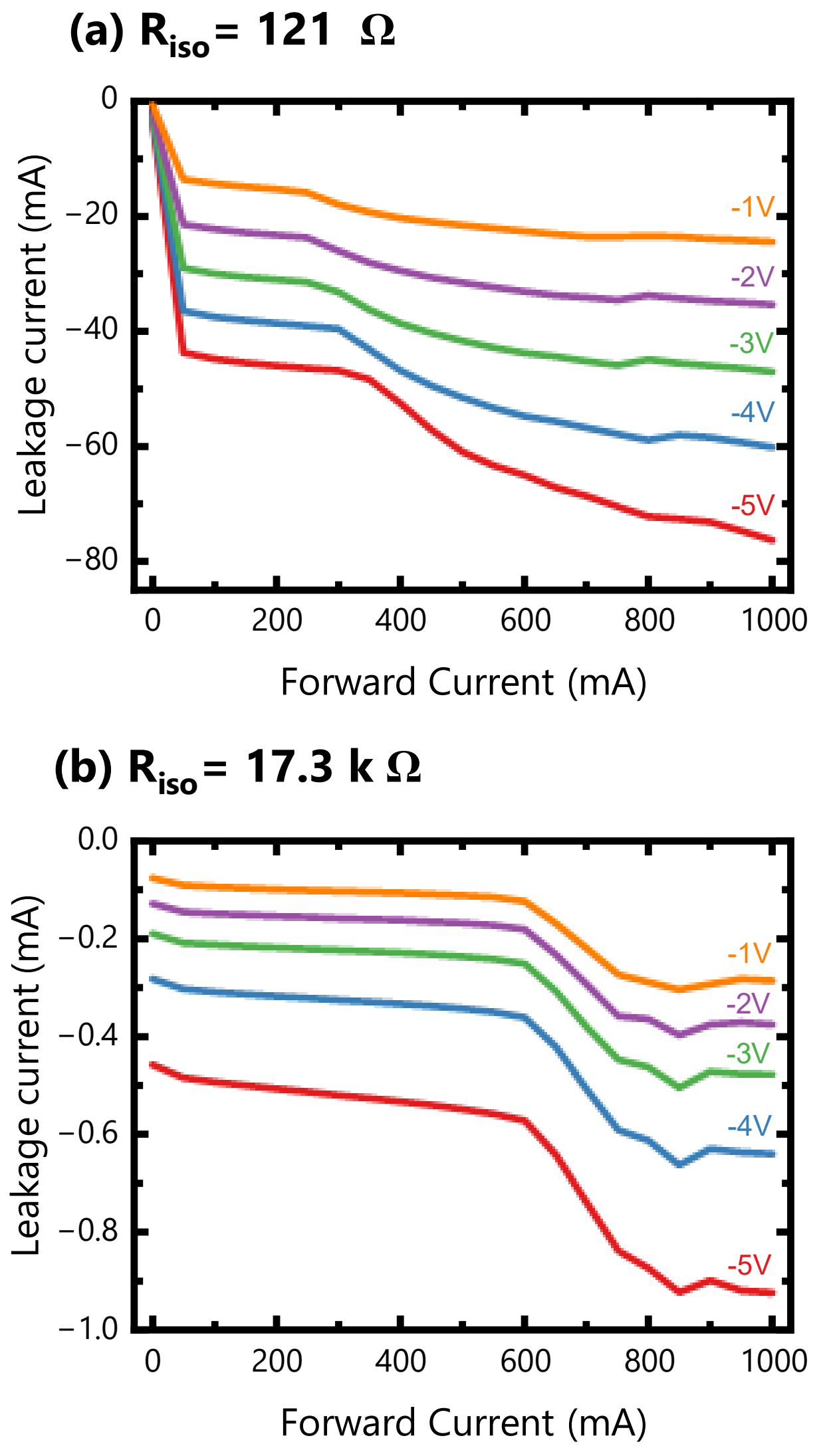

**Fig. 4.** Measured current in the isolation section vs. forward biasing at different absorber voltages for the (a) baseline device A and (b) optimized device D.

The PIV characteristic of device operation as CW laser diodes was measured in the same biasing configuration using a Thorlabs PM400 photodiode-integrating sphere and is shown in **Fig. S4(a)** (supplementary material) for the co-designed device. The lasing threshold current remained consistently at ~100 mA for the two extremes of absorber reverse bias. The slope efficiency dropped from 0.021 W/A with the absorber left unbiased to 0.013 W/A at 5 V reverse bias as a result of an increase in inter-band and exciton absorption. No sign of roll-over was observed for driving currents up to 800 mA and reverse bias of 5 V. The optical spectrum was obtained using a Bristol 771 series laser spectrum analyzer (4 GHz resolution), with a single mode fiber coupled to the facet of the laser as shown in **Fig. S4(b)** (supplementary material), confirming single mode lasing**.** There was a small (~0.3 nm) red shift of the peak by about 3 dB when the absorber was biased, however the effect was not so large as it would be in the case of stronger device heating which could be caused by high leakage currents.

**Fig. 5** shows the comparison of the device variants in the electrical and optical performance space, reporting the resistance (kΩ) of the two-section isolation and the lasing power (mW) at some fixed bias. The inability of device C to lase, despite achieving the best $R_{iso}$~40 kΩ, illustrates the importance of waveguide tapering in preserving low-loss propagation in the laser cavity with additional etch features. At the other end, the baseline device A operated successfully as a CW laser with ~9 mW optical output but its poor electrical isolation ($R_{iso}$=121 Ω) points to compromised performance at higher absorber reverse biases. The impact of the leakage prevention etch on electrical isolation was underscored by a 5X higher $R_{iso}$ in device B, although optical performance was negatively affected as expected. In contrast to the extremes of devices A and C, the tapered device D strikes the balance in the co-design of both the electrical and optical performance space with $R_{iso}$=17.3 kΩ and ~8 mW lasing. The tapered device with 2.5X $L_{iso}$ (type E) demonstrated a slightly better isolation ($R_{iso}$= 22.13 kΩ) at the expense of ~5X lower lasing power which can be attributed to higher propagation losses in the longer deeply etched section.

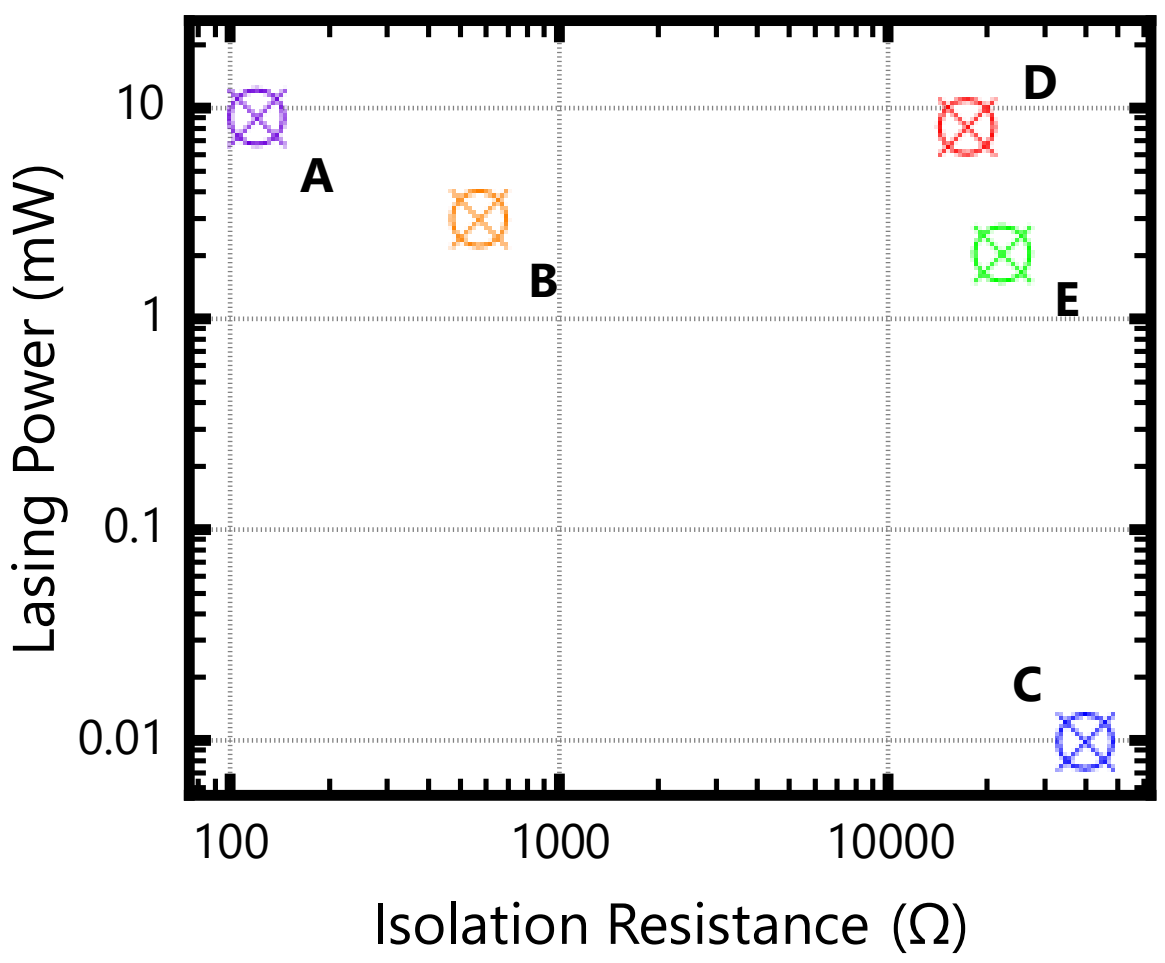

**Fig. 5.** Comparison of isolation resistance and lasing power at a fixed bias for all device variants.

In conclusion, it was shown that achieving high electrical isolation along the longitudinal direction of RWGs in GaSb-based MQW laser diodes demands deeply etched features in light of the inherent problem of lateral current leakage originating from high conductivity in p-doped AlGaAsSb cladding layers. We experimentally evaluated the independent impact of different etching features on inter-section resistance and leveraged a simulation based co-design process to mitigate optical losses and reflection with a ridge-to-strip waveguide taper in the isolation section. A maximum inter-section resistance of ~40 kΩ was achieved in deep mesa etched and non-tapered device variant. The optimized device functioned in the single mode lasing regime with a CW power of ~8 mW and sub-100 mA threshold, while achieving an isolation resistance of 17.3 kΩ, which is about 17 times better isolation compared to previously reported two-section GaSb devices.  Besides CW laser diodes, this design approach can be applied to devices facing similar isolation problems or requiring varying etch depths along the waveguide, such as electro-absorption modulators combined with semiconductor optical amplifiers, mode locked laser diodes, or high-speed photodiodes. Moreover, the design can be utilized as a building block towards monolithic GaSb-based PIC platforms as an enabler of scalable gas and biomedical sensor applications in the MIDIR spectral range.

## Supplementary Material

See the supplementary material for the waveguide taper optical simulations, the epitaxial fabrication, the electrical isolation study in strip waveguides, as well as the PIV characterization and optical spectrum measurements.

## Author Contributions

**Md Ajwaad Zaman Quashef:** Data curation (lead); Formal analysis (lead); Investigation (lead); Methodology (supporting); Software (equal); Visualization (lead); Validation (lead); Writing – original draft (lead); Writing – review & editing (lead). **Nouman Zia:** Conceptualization (lead); Formal analysis (supporting); Software (equal); Methodology (lead). **Jukka Viheriälä:** Formal analysis (equal); Funding acquisition (supporting); Project administration (equal); Resources (lead); Supervision (equal); Writing – review & editing (supporting). **Mircea Guina:** Conceptualization (supporting); Funding acquisition (lead); Investigation (supporting); Supervision (lead); Writing – review & editing (equal).

## Acknowledgements

This work was supported by Academy of Finland PoC project IntegrateQT (decision number 359431) and the Finnish Ministry of Education and Culture through the PREIN/I-DEEP doctoral pilot (VN/3137/2024-OKM-4). We also acknowledge the Flagship for Photonics Research and Innovation (PREIN) with decision number Research Council of Finland/2024/368650. We acknowledge the device fabrication support received from Dr. Heikki Virtanen and Dr. Eero Koivusalo.

## Data Availability Statement

The data that support the findings of this study are available from the corresponding author upon reasonable request.